\documentclass[prb,twocolumn,showpacs,superscriptaddress]{revtex4}
\usepackage{amsmath,amsfonts,amssymb,amsbsy,graphicx}

\renewcommand{\epsilon}{\varepsilon}
\renewcommand{\vec}[1]{\boldsymbol{#1}}
\DeclareMathOperator{\K}{K}
\newcommand{\Zb}{\mathbb{Z}}
\newcommand{\bz}{\mathrm{BZ}}

\begin{document}
\title{Harmonic Stability Analysis of the 2D Square and Hexagonal Bravais Lattices\\
  for a Finite--Ranged Repulsive Pair Potential.\\
  Consequence for a 2D System of Ultracold Composite Bosons.}
\author{D.J. Papoular} \email{david.papoular@lptms.u-psud.fr}
\affiliation{Laboratoire de Physique Th\'eorique et Mod\`eles
  Statistiques, Universit\'e Paris--Sud, 91405 Orsay, France}
\affiliation{\'Ecole Normale Sup\'erieure, Paris, France}
\date{\today}

\begin{abstract}
  We consider a classical, two--dimensional system of identical
  particles which interact \textit{via} a finite--ranged, repulsive
  pair potential. We assume that the system is in a crystalline
  phase. We calculate the normal vibrational modes of a
  two--dimensional square Bravais lattice, first analytically within
  the nearest--neighbour approximation, and then numerically, relaxing
  the preceding hypothesis. We show that, in the harmonic
  approximation, the excitation of a transverse vibrational mode leads
  to the breakdown of the square lattice. We next study the case of
  the hexagonal Bravais lattice and we show that it can be stable with
  respect to lattice vibrations. We give a criterion determining
  whether or not it is stable in the nearest--neighbour
  approximation. Finally, we apply our results to a two--dimensional
  system of composite bosons and infer that the crystalline phase of
  such a system, if it exists, corresponds to a hexagonal lattice.
\end{abstract}

\pacs{63.22.-m, 64.70.dg, 03.75.Ss}
\maketitle

\section*{Introduction}
Recent developments in atomic Physics, spanning the past decade, have
made it possible to observe states of matter which, so far, had only
been considered from a theoretical point of view. Bose--Einstein
condensation, experimentally achieved in
1995\cite{anderson:1995,davis:1995}, is a landmark among these
triumphs. More recently, much experimental effort has been directed
towards the study of ultracold Fermi gases, allowing for instance an
experimental study of the BEC---BCS crossover
regime\cite{holland:2001}.

A stimulating new prospect for atomic physicists is the study of
ultracold degenerate Fermi gases containing two \textsl{different}
types of atomic species, such as ${}^6\mathrm{Li}$ and
${}^{40}\mathrm{K}$.  In such a gas, it is possible to tune the
strength of the interaction between the two species by varying the
interspecies scattering length using a Feshbach
resonance\cite{kohler:2006}. For a positive scattering length, bosonic
dimers appear, containing one atom of each type\cite{petrov:2004}.
Being in their highest--energy bound states, these composite bosons
are stable with respect to collisional relaxation into deeper--bound
states. Two--component degenerate Fermi gases have recently been
obtained experimentally\cite{taglieber:2008}, and interspecies
Feshbach resonances have been identified\cite{wille:2008}.

If the difference between the two atomic masses is large enough,
composite bosons interact \textit{via} an effective pair potential
which is repulsive. An expression for this pair potential has been
derived in the Born--Oppenheimer approximation\cite{petrov:2007}. In
the quasi--two--dimensional regime, this system has been shown to
exhibit a crystalline phase for suitable values of the density and of
the atomic mass ratio\cite{petrov:2007}.

In this work, we perform an analysis of the stability of the
two--dimensional Bravais lattices with high symmetry properties,
namely the square lattice and the hexagonal lattice, with respect to
classical harmonic vibrations. We use the nearest--neighbour
approximation to derive analytic expressions for the dispersion
relations of both lattices. We show that, in the case of a
finite--ranged repulsive pair potential, the square lattice is
unstable for all values of the density. Still in the
nearest--neighbour approximation, we give a criterion determining the
range of densities for which the hexagonal lattice is stable. In both
cases, we present numerical results which show that taking into
account more rings of neighbours does not affect the qualitative
behaviour of the system. Applying our results to a two--dimensional
system of ultracold composite bosons, we infer that, for values of the
mass ratio and density leading to a crystalline phase, the system
crystallises into a hexagonal lattice.

\section{\label{sec:harm_theory} Harmonic theory of lattice vibrations
  for particles interacting \textit{via} a pair potential}

For the sake of clarity, we first recall the general
method\cite{ashcroft:1976} for the calculation of the normal
vibrational modes of a two--dimensional crystalline solid.

We consider a two--dimensional system of identical particles which we
describe within the framework of classical mechanics.  We assume that
these particles interact only \textit{via} a finite--ranged pair
potential $U(R)$, where $R$ is the distance between two particles. We
also assume that the system is in a crystalline phase corresponding to
a two--dimensional Bravais lattice generated by two vectors
$\vec{a}_1$ and $\vec{a}_2$: at classical equilibrium, there is a
particle at each lattice site $\vec{r}_p$ such that
\begin{equation}\label{eq:lattice_site}
  \vec{r}_p= p_1 \vec{a}_1 + p_2 \vec{a}_2,
\end{equation}
where $p=(p_1,p_2)$ is a pair of integers.

In order to characterise the stability of such a lattice, we shall
study the propagation of lattice waves. For that purpose we shall
first write the Lagrangian of the system in the harmonic
approximation, then derive from it the equations of motion, and
finally look for plane--wave solutions to these equations.  The
lattice is stable if the frequencies of the normal modes thus found
are all real; on the contrary, \textit{i.e.} if there is at least
one normal mode
with an imaginary frequency, the lattice is unstable.

\subsection{Lagrangian of the crystal
                       \label{sec:harm_lagrangian}}
The total potential energy of the crystal is
\begin{equation}
  U^\mathrm{tot}((\vec{u}_n))=
  \frac{1}{2} \sum_{p\neq q}
  U\left( |\vec{r}_p+\vec{u}_p-\vec{r}_q-\vec{u}_q| \right),
\end{equation}
where $\vec{u}_n$ is the displacement of atom $n$ from the
corresponding lattice site $\vec{r}_n$, and the double sum over $p$
and $q$ spans all pairs of lattice sites.  We expand $U^\mathrm{tot}$
up to second order in the displacements $\vec{u}_n$. The constant term
has no incidence on the dynamics of the crystal lattice and will
therefore be dropped in subsequent calculations. The linear term
cancels out when the double sum is performed.
Consequently, the harmonic approximation to
$U^\mathrm{tot}$ is a quadratic function of the $((\vec{u}_n))$:
\begin{equation}
  U^\mathrm{tot}_\mathrm{harm}((\vec{u}_n))=
  \frac{1}{2} \sum_{pq}
  {}^t \vec{u}_p \Lambda_{pq} \vec{u}_q,
\end{equation}
where the real--space dynamical matrices $\Lambda_{pq}$ are real
$2\times 2$ matrices defined by\cite{ashcroft:1976}:
\begin{equation}\label{eq:dynmat_real}
  \Lambda_{pq}^{ij}=
  \left.
    \frac{\partial^2 U^\mathrm{tot}((\vec{u}_n))}
    {\partial u_p^i \, \partial u_q^j}
  \right|_{(\vec{u}_n=\vec{0})}.
\end{equation}

In the harmonic approximation, the Lagrangian of the crystal is thus:
\begin{equation}\label{eq:harm_lagrangian}
  L= \frac{1}{2} m \sum_p \dot{\vec{u}_p}^2
  -
  \frac{1}{2} \sum_{pq} 
  {}^t \vec{u}_p \Lambda_{pq} \vec{u}_q,
\end{equation}
where the first term is the total kinetic energy of the system and the
second term is the harmonic approximation to the total potential
energy. $m$ is the mass of each particle in the system.

  The $\Lambda_{pq}^{ij}$'s are endowed with well--documented
properties\cite{wallace:1972,ashcroft:1976},
among which tensor symmetry, invariance under spatial inversion,
and invariance under lattice translations. Additionally, the
following expression, valid if only pairwise interactions are
considered, greatly simplifies their evaluation:
\begin{equation}\label{eq:dynmat_real_expr}
  \Lambda_{0p}^{ij}=-\left.
    \frac{\partial^2 U(|\vec{r}_p + \vec{u}|)}
         {\partial u_i \, \partial u_j}
  \right|_{\vec{u}=\vec{0}}.
\end{equation}

We now introduce the momentum--space dynamical
matrix $\Lambda(\vec{k})$, defined as the
discrete Fourier transform of the $\Lambda_{0p}$
over the crystal lattice:
\begin{equation}\label{eq:dynmat_mom}
  \Lambda(\vec{k})=\sum_p \Lambda_{0p}\,
  e^{i\vec{k}\cdot\vec{r}_p}
  =-2\sum_p
  \Lambda_{0p} \sin^2\left(\frac{1}{2}\vec{k}\cdot\vec{r}_p\right).
\end{equation}
For a given
wavevector $\vec{k}$, $\Lambda(\vec{k})$ is a
real symmetric matrix. As such, it has two real
orthogonal eigenvectors
$\vec{\epsilon}_{(\vec{k},1)}$ and
$\vec{\epsilon}_{(\vec{k},2)}$.

\subsection{Equations of motion}
Using the translational invariance property of
the $\Lambda_{pq}$'s, the (classical) equation
of motion for atom $n$, resulting from the Lagrangian
\ref{eq:harm_lagrangian}, reads:
\begin{equation}\label{eq:harm_motion}
  m\ddot{\vec{u}_n}=
  -\sum_{p} \Lambda_{0p} \vec{u}_{n+p}.
\end{equation}
In the harmonic approximation, the motion of the
particles in the crystal is thus determined by a
set of $N$ coupled linear equations similar to
Equation \ref{eq:harm_motion}, where $N$ is the
number of independent particles in the system.

We now determine the normal vibrational modes of
the system, \textit{i.e.} we look for a
plane--wave solution to the equations of motion :
\begin{equation}\label{eq:planewave}
  \vec{u}_n= A\,\vec{\epsilon}
  \exp i \left(
    \vec{k}\cdot\vec{r}_n - \omega t
  \right),
\end{equation} 
where $\vec{\epsilon}$ is the polarisation of the
mode, $\vec{k}$ is its wavevector, and
$\frac{\omega}{2\pi}$ is its frequency.  $A$ is
an arbitrary complex number characterising the
amplitude and global phase of the collective
vibrational motion. Inserting Equation
\ref{eq:planewave} into Equation
\ref{eq:harm_motion}, we obtain:
\begin{equation}\label{eq:dynmat_eigen}
   m\omega^2\,\vec{\epsilon} = \Lambda(\vec{k})\,\vec{\epsilon}.
\end{equation}
where $\Lambda(\vec{k})$ is the momentum--space
dynamical matrix defined in Section
\ref{sec:harm_lagrangian}.  Equation
\ref{eq:dynmat_eigen} shows that for a given
wavevector $\vec{k}$, there are two possible
polarisations $\vec{\epsilon}_{(\vec{k},1)}$ and
$\vec{\epsilon}_{(\vec{k},2)}$ which are the two
eigenvectors of the dynamical matrix
$\Lambda(\vec{k})$. The corresponding eigenvalues
$m\omega_1^2(\vec{k})$ and $m\omega_2^2(\vec{k})$
yield their respectives frequencies
$\omega_1(\vec{k})$ and $\omega_2(\vec{k})$.

In the harmonic approximation, the classical
dynamical properties of the crystal are thus
completely determined by the dynamical matrices
$\Lambda(\vec{k})$.

\section{The specific case of the 2D square lattice for a purely
  repulsive pair potential\label{sec:square}}
We now apply the formalism summarised in Section \ref{sec:harm_theory}
to the specific case of the square Bravais lattice, generated by two
vectors $\vec{a}_1$ and $\vec{a}_2$ such that:
\begin{equation}
  |\vec{a}_1|=|\vec{a}_2|=d
  \quad \text{and} \quad
  (\widehat{\vec{a}_1,\vec{a}_2})=\frac{\pi}{2}.
\end{equation}

The wavevectors $\vec{k}$ are most conveniently described in the
reciprocal lattice basis $(\vec{a}_1^*,\vec{a}_2^*)$ defined by
$\vec{a}_i^* \cdot \vec{a}_j = 2\pi\cdot\delta_{ij}$. The reciprocal
lattice of a square lattice is also a square lattice:
\begin{equation}
  |\vec{a}_1^*|=|\vec{a}_2^*|=\frac{2\pi}{d}
  \quad \text{and} \quad
  (\widehat{\vec{a}_1^*,\vec{a}_2^*})=\frac{\pi}{2}.
\end{equation}

\subsection{Analytical expression for the dispersion relation in the
                        nearest--neighbour approximation}
                      We first derive the expression for
                      $\Lambda_{0p}$, where the lattice index
                      $p=(p_1,p_2)\in\Zb^2$, using Equation
                      \ref{eq:dynmat_real_expr}:
                      \begin{widetext}
                        \begin{equation}
                          \Lambda_{0p}=-\frac{d^2}{r_p^2}
                          \begin{bmatrix}
                            p_1^2\,U''(r_p) + p_2^2\,\frac{U'(r_p)}{r_p}            & p_1 p_2\,\left( U''(r_p) - \frac{U'(r_p)}{r_p} \right) \\
                            p_1 p_2\,\left( U''(r_p) -
                              \frac{U'(r_p)}{r_p} \right ) &
                            p_2^2\,U''(r_p)
                            +p_1^2\,\frac{U'(r_p)}{r_p}
                          \end{bmatrix}.
                        \end{equation}
                      \end{widetext}

                      Next, the momentum--space dynamical matrix
                      $\Lambda(\vec{k})$ can be calculated from
                      Equation \ref{eq:dynmat_mom}. An exact
                      calculation of $\Lambda(\vec{k})$ would require
                      calculating an infinite series spanning all
                      sites of the two--dimensional Bravais
                      lattice. However, assuming that the range of the
                      pair potential $U(R)$ is small compared to the
                      lattice spacing $d$, the nearest--neighbour
                      approximation can be used. The right--hand side
                      of Equation \ref{eq:dynmat_mom} then reduces to
                      a sum of five terms, corresponding to the
                      reference lattice site $p=(0,0)$ and to its four
                      nearest neighbours. Letting $\vec{k}=k_1
                      \vec{a}_1^* +k_2 \vec{a}_2^*$, we thus obtain
                      the following expression for $\Lambda(\vec{k})$:
                      \begin{widetext}
                        \begin{equation}
                          \Lambda(\vec{k})=4
                          \begin{bmatrix}
                            U''(d)\sin^2(\pi k_1) + \frac{U'(d)}{d}\sin^2(\pi k_2)  & 0\\
                            0 & \frac{U'(d)}{d}\sin^2(\pi k_1) +
                            U''(d)\sin^2(\pi k_2)
                          \end{bmatrix}.
                        \end{equation}
                      \end{widetext}

                      In the nearest--neighbour approximation,
                      $\Lambda(\vec{k})$ is a diagonal
                      matrix. According to the results of Section
                      \ref{sec:harm_theory}, the analytical
                      expressions for the two (acoustic) branches of
                      the dispersion relation can be read off the
                      diagonal elements of $\Lambda(\vec{k})$:
                      \begin{equation}\label{eq:disprel_square}
                        \begin{cases}
                          m\omega_1^2(\vec{k})=& 4 \left(
                            U''(d)\sin^2(\pi k_1) +
                            \frac{U'(d)}{d}\sin^2(\pi k_2)
                          \right)\\
                          m\omega_2^2(\vec{k})=& 4 \left(
                            \frac{U'(d)}{d}\sin^2(\pi k_1) +
                            U''(d)\sin^2(\pi k_2)
                          \right)\\
                        \end{cases}
                      \end{equation}
                      Because the dynamical matrix is diagonal, the
                      allowed polarisations depend only on the branch
                      of the dispersion relation that is considered
                      (they do not depend on the wavevector). The
                      first branch --- $\omega_1^2(\vec{k})$ ---
                      corresponds to the polarisation
                      $\vec{\epsilon}_1=\frac{\vec{a}_1}{|\vec{a}_1|}$,
                      whereas the second branch ---
                      $\omega_2^2(\vec{k})$ --- corresponds to the
                      polarisation
                      $\vec{\epsilon}_2=\frac{\vec{a}_2}{|\vec{a}_2|}$.

                      Equation \ref{eq:disprel_square} is compatible
                      with the four--fold symmetry of the
                      two--dimensional square lattice. Indeed, let
                      $\vec{k'}$ be the image of $\vec{k}$ under the
                      vector rotation of angle~$\frac{\pi}{2}$:
                      $\vec{k'}=-k_2\vec{a}_1^*+k_1\vec{a_2}^*$.
                      Equation \ref{eq:disprel_square} yields
                      $\omega_1^2(\vec{k'})=\omega_2^2(\vec{k})$ and
                      $\omega_2^2(\vec{k'})=\omega_1^2(\vec{k})$.

\subsection{\label{sec:square_lattice_unstable}
  Instability of the square lattice for a purely repulsive
  pair potential}
In the harmonic approximation, a crystal lattice is stable if lattice
waves can propagate through the crystal for all wavevectors $\vec{k}$
in the first Brillouin zone of the lattice. We now show that this is
not the case for the square lattice if the pair potential is purely
repulsive.

\paragraph*{Nearest--neighbour approximation.}
Let us consider a wavevector lying along $\vec{a}_1^*$: $\vec{k}=k_1
\vec{a}_1^*$ . Equations \ref{eq:disprel_square} reduce to:
\begin{equation}\label{eq:disprel_square_unstable}
  \begin{cases}
    m\omega_1^2(\vec{k})=&
    4 U''(d)\sin^2(\pi k_1)\\
    m\omega_2^2(\vec{k})=& 4 \frac{U'(d)}{d}\sin^2(\pi k_1)
  \end{cases}
\end{equation}
For a purely repulsive potential, $U'(d)<0$ for all possible values of
the lattice spacing $d$. Consequently, $\omega_2^2$ is negative, and
therefore the frequency of the normal mode with wavevector
$\vec{k}=k_1\vec{a}_1^*$ and polarisation $\vec{\epsilon}_2$ is not
defined.  Physically, this means that the propagation of a
\textsl{transverse} normal mode (\textit{i.e.} a normal mode with
$\vec{k}\perp\vec{\epsilon}$) with a wavevector along $\vec{a}_1^*$
would \textsl{break} the crystal lattice. Because of the four--fold
symmetry of the square lattice, the same results and conclusions are
valid for a transverse mode with a wavevector along $\vec{a}_2^*$.
 
Consequently, in the particular case of a purely repulsive pair
potential, the two--dimensional square lattice is \textsl{not stable}.

\begin{figure}
  \begin{center}
    \includegraphics[width=\linewidth]{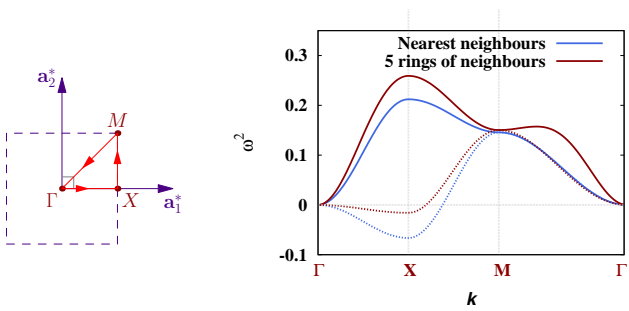}
    \caption{\label{fig:disprel_square} Left: Brillouin zone of the
      square Bravais lattice, with the high--symmetry points
      $\Gamma(0,0)$, $X(\frac{1}{2},0)$, and
      $M(\frac{1}{2},\frac{1}{2})$. Right: the two branches
      $\omega_1^2(\vec{k})$ (solid lines) and $\omega_2^2(\vec{k})$
      (dashed lines) of the dispersion relation of a two--dimensional
      square lattice, for wavevectors $\vec{k}$ with origin $\Gamma$
      and whose tips lie on the $\Gamma$--$X$--$M$--$\Gamma$ path
      represented in red on the diagram on the left.  The pair
      potential is the one characterising the interaction of two
      composite bosons in the two--dimensional case. The lattice
      spacing is $d=2.0$ in units of the composite--boson molecular
      size.  The total mass $m$ of the composite bosons is taken to be
      unity.  The blue graphs correspond to the analytical result in
      the nearest--neighbour approximation; the red graphs are
      numerical results taking into account five rings of neighbours.
      Note the lifting of the $\omega_{1,2}^2(\vec{k})$ degeneracy
      along the $M$--$\Gamma$ branch as soon as next--nearest neighbours
      are taken into account.}
  \end{center}
\end{figure}

The variations of $\omega_1^2(\vec{k})$ and $\omega_2^2(\vec{k})$, in
the nearest--neighbour approximation (\textit{i.e.} as given by the
analytical expressions \ref{eq:disprel_square_unstable}), are
represented in Figure \ref{fig:disprel_square} for wavevectors
$\vec{k}$ whose tips lie on the high--symmetry axes of the Brillouin
zone\cite{kittel:1987}, in the case of the pair potential
characterising the two--dimensional interactions of composite bosons
at temperature $T=0\,\mathrm{K}$ \cite{petrov:2007}. The branch
$\omega_2^2(\vec{k})$ is negative for all wavevectors along
$\vec{a}_1^*$ ($\Gamma$--$X$ part of the plot of $\omega^2(\vec{k})$).

\paragraph*{Numerical results including more distant neighbours.}
In order to go beyond the nearest--neighbour approximation, we have
performed numerical calculations including more distant
neighbours. For that purpose we have written a Python program which
evaluates the lattice sums involved in Equation
\ref{eq:dynmat_mom} for a finite--sized square lattice with 100
particles in both the $\vec{a}_1$ and $\vec{a}_2$ directions. The pair
potential is finite--ranged, and the numerical results for
$\omega_1^2(\vec{k})$ and $\omega_2^2(\vec{k})$ therefore converge
quickly as a function of the radius of the disk of neighbours taken
into account.  The results of these calculations are represented in
Figure \ref{fig:disprel_square}. The branch $\omega_2^2(\vec{k})$
remains negative for $\vec{k}$ vectors along $\vec{a}_1^*$.
Consequently, the effect described in the preceding paragraph is not
an artefact due to the nearest--neighbour approximation: \textsl{in
  the particular case of a purely repulsive pair potential, and in the
  harmonic approximation, the two--dimensional square lattice is
  unstable}.

\paragraph*{Geometrical interpretation of the instability.}
\begin{figure}[b]
  \begin{center}
    \includegraphics[width=\linewidth]{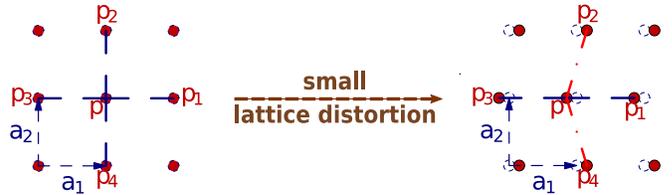}
    \caption{\label{fig:square_inst} Distortion of the crystal lattice
      due to the transverse vibrational mode with polarisation
      $\vec{\epsilon}_1$ and wavevector
      $\vec{k}=\frac{1}{2}\vec{a}_2^*$.}
  \end{center}
\end{figure}
Let us consider the transverse vibrational mode with polarisation
$\vec{\epsilon}_1$ and wavevector $\frac{1}{2}\vec{a}_2^*$. Equation
\ref{eq:planewave} shows that the displacement of atom $n=(n_1,n_2)$
from the corresponding lattice site is (omitting the time dependence
$e^{i\omega_1 t})$:
\begin{equation}
  \vec{u}_n=A\,\vec{\epsilon}_1\cdot (-1)^{n_2}.
\end{equation}
We shall consider the case of a small amplitude $A$.  The positions of
the particles for a lattice which is at rest on the one hand, and when
the mode described above is excited on the other hand, are compared on
Figure \ref{fig:square_inst}. In the absence of vibrations, the
distance of atom $p$ to its four nearest neighbours $p_1$, $p_2$,
$p_3$, and $p_4$ is exactly equal to the lattice spacing $d$
(left--hand side of the figure).  The mode described above modifies
the distances between atom $p$ and some of its neighbours. The
distance between $p$ and $p_1$ remains unchanged (\textit{i.e.} equal
to $d$), as well as the one between $p$ and $p_3$; however, the
distance between $p$ and $p_2$, as well as the one between $p$ and
$p_4$, are increased to $d\left(1+\frac{A^2}{d^2}\right)$.  All four
distances are increased (or remain unchanged), and the pair potential
is repulsive, therefore the total interaction energy between atom $p$
and its four nearest neighbours is \textsl{decreased}.  The slightly
distorted lattice represented on the right--hand side of Figure
\ref{fig:square_inst} therefore has a lower potential energy than the
square lattice represented on the left--hand side, which entails that
the square lattice is \textsl{not} a stable equilibrium position.

\section{\label{sec:hex} The case of the 2D Hexagonal Bravais lattice}
In Section \ref{sec:square}, we showed that, within the framework of
the harmonic approximation for crystal vibrations, the
two--dimensional square Bravais lattice is not stable. In the present
section, we apply the same formalism to the two--dimensional hexagonal
lattice and show that, in contrast with the former, the latter is
stable in the harmonic approximation.

The two--dimensional hexagonal Bravais lattice is generated by two
vectors $\vec{a}_1$ and $\vec{a}_2$ such that:
\begin{equation}
  |\vec{a}_1|=|\vec{a}_2|=d
  \quad \text{and} \quad
  (\widehat{\vec{a}_1,\vec{a}_2})=\frac{2\pi}{3}.
\end{equation}
As for the square lattice, we introduce the reciprocal lattice basis
$(\vec{a}_1^*,\vec{a}_2^*)$, defined as before by
$\vec{a}_i^*\cdot\vec{a}_j=\delta_{ij}$. The reciprocal lattice of a
hexagonal lattice is also a hexagonal lattice:
\begin{equation}
  |\vec{a}_1^*|=|\vec{a}_2^*|=\frac{2\pi}{d}\frac{2}{\sqrt{3}}
  \quad \text{and} \quad
  (\widehat{\vec{a}_1^*,\vec{a}_2^*})=\frac{\pi}{3}.
\end{equation}

\paragraph*{Dispersion relation.}
Equation \ref{eq:dynmat_real_expr} yields the following expression for
$\Lambda_{0p}$, where $p=(p_1,p_2)\in\Zb^2$:
\begin{widetext}
  \begin{equation}
    \Lambda_{0p}=-\frac{1}{4}\frac{d^2}{r_p^2} 
    \begin{bmatrix}
      (2p_1-p_2)^2\,U''(r_p) +3p_2^2\,\frac{U'(r_p)}{r_p}&
      \sqrt{3}\,p_2(2p_1-p_2)\,(U''(r_p)-\frac{U'(r_p)}{r_p})\\
      \sqrt{3}\,p_2(2p_1-p_2)\,(U''(r_p)-\frac{U'(r_p)}{r_p})&
      3p_2^2\,U''(r_p)+(2p_1-p_2)^2\frac{U'(r_p)}{r_p}\\
    \end{bmatrix}.
  \end{equation}
  Using Equation \ref{eq:dynmat_mom} in the nearest--neighbour
  approximation, we then derive the expression for $\Lambda(\vec{k})$:
  \begin{equation}\label{eq:dynmat_hex}
    \Lambda(\vec{k})=
    \begin{bmatrix}
      4U''(d) s_1^2+(U''(d)+3\frac{U'(d)}{d})(s_2^2+s_3^2)&
      \sqrt{3}(U''(d)-\frac{U'(d)}{d})(s_3^2-s_2^2)\\
      \sqrt{3}(U''(d)-\frac{U'(d)}{d})(s_3^2-s_2^2)&
      4\frac{U'(d)}{d}+(3U''(d)+\frac{U'(d)}{d})(s_2^2+s_3^2)
    \end{bmatrix}.
  \end{equation}
\end{widetext}
where $s_1=\sin\left(\frac{1}{2}\vec{k}\cdot\vec{a}_1\right)$,
$s_2=\sin\left(\frac{1}{2}\vec{k}\cdot\vec{a}_2\right)$, and
$s_3=\sin\left(\frac{1}{2}\vec{k}\cdot\left(\vec{a}_1+\vec{a}_2\right)\right)$.
Equation \ref{eq:dynmat_hex} yields the following approximate
analytical expression for the two branches of the dispersion relation,
which are obtained as the two eigenvalues of $\Lambda(\vec{k})$:
\begin{equation}\label{eq:disprel_hex}
  \begin{aligned}
    m\omega_{1,2}^2(\vec{k})&=
    2\left(U''(d)+\frac{U'(d)}{d}\right)(s_1^2+s_2^2+s_3^2)\\
    &\pm 2\left(U''(d)-\frac{U'(d)}{d}\right) s_0^2,
  \end{aligned}
\end{equation}
where $s_0^2=\sqrt{(s_1^2+s_2^2+s_3^2)^2-3(s_1^2 s_2^2 + s_2^2 s_3^3 +
  s_3^2 s_1^2)}$.  Equation \ref{eq:disprel_hex} is symmetrical in
$s_1$, $s_2$, and $s_3$, and is thus compatible with the six--fold
symmetry of the two--dimensional hexagonal lattice.

The polarisations corresponding to $\omega_{1,2}$ are
$\vec{\epsilon}_{1,2}=\epsilon^x_{1,2}\vec{e}_x+\epsilon^y_{1,2}\vec{e}_y$,
where $(\vec{e}_x,\vec{e}_y)$ is the two--dimensional direct
orthonormal basis with $\vec{e}_x$ along $\vec{a}_1$, and
\begin{equation}\label{eq:hex_pol}
  \begin{cases}
    {\epsilon^x_{1,2}}^2(\vec{k})=
    \frac{3(s_3^2-s_2^2)^2}{3(s_3^2-s_2^2)^2+(2s_1^2-s_2^2-s_3^2\mp 2s_0^2)^2}\\
    {\epsilon^y_{1,2}}^2(\vec{k})= \frac{(2s_1^2-s_2^2-s_3^2\mp
      2s_0^2)^2}{3(s_3^2-s_2^2)^2+(2s_1^2-s_2^2-s_3^2\mp 2s_0^2)^2}
  \end{cases}
\end{equation}
The frequencies $\omega_{1,2}(\vec{k})$ of the normal vibrational
modes depend on the first and second derivatives of the pair
potential, whereas the corresponding polarisations are independent of
the particular shape $U(x)$ of this potential. However, in contrast to
the case of the square lattice, the polarisations
$\vec{\epsilon}_{1,2}(\vec{k})$ for the hexagonal lattice do depend on
the considered wavevector $\vec{k}$.

\begin{figure}
  \begin{center}
    \includegraphics[width=\linewidth]{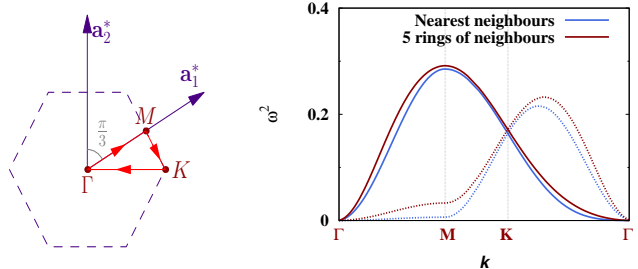}
    \caption{\label{fig:disprel_hex} Left: Brillouin zone of the
      two--dimensional hexagonal Bravais lattice, with the
      high--symmetry points $\Gamma(0,0)$, $M(\frac{1}{2},0)$, and
      $K(\frac{1}{3},\frac{1}{3})$. Right: the two branches
      $\omega_{1,2}(\vec{k})^2$ of the dispersion relation for the
      hexagonal lattice, for wavevectors $\vec{k}$ with origin
      $\Gamma$ and whose tips lie on the $\Gamma$--$M$--$K$--$\Gamma$
      path represented in red on the diagram on the left. The pair
      potential is the one characterising the interaction of two
      composite bosons in the two--dimensional case.  The blue graphs
      correspond to the analytical result in the nearest--neighbour
      approximation; the red graphs are numerical results taking into
      account five rings of neighbours.  As in Figure
      \ref{fig:disprel_square}, the lattice spacing is $d=2.0$ and the
      total mass is $m=1$, in the same units.}
  \end{center}
\end{figure}

The variations of $\omega_{1,2}^2(\vec{k})$ are represented in Figure
\ref{fig:disprel_hex} in the case of the pair potential characterising
the interaction of two composite bosons in the fully two--dimensional
situation\cite{petrov:2007}, for wavevectors $\vec{k}$ whose tips lie
on the high--symmetry axes of the Brillouin zone\cite{kittel:1987}.
The analytical results obtained in the nearest--neighbour approximation
(Equation \ref{eq:disprel_hex}) are compared to numerical calculations
taking into account five rings of neighbours on a finite--sized
hexagonal lattice with 100 independent particles in both the
$\vec{a}_1$ and $\vec{a}_2$ directions.  Both calculations have been
performed for the lattice parameter $d=2.0$ (in units of the
composite--boson molecular size).  In both cases, the two branches
$\omega_{1,2}^2(\vec{k})$ of the dispersion relation are positive for
all wavevectors $\vec{k}$ in the Brillouin zone.  Consequently,
contrary to the results presented in Section
\ref{sec:square_lattice_unstable} for the square Bravais lattice,
there is a range of densities $\rho$ for which the two--dimensional
hexagonal Bravais lattice of composite bosons is stable with respect
to harmonic lattice vibrations.

\paragraph*{Stability criterion for the hexagonal lattice.}
We now derive, in the nearest--neighbour approximation, a simple
criterion on the relative values of $U''(d)$ and $U'(d)/d$ determining
whether the hexagonal lattice is stable or not.

A two--dimensional hexagonal Bravais lattice with a given lattice
spacing $d$ (\textit{i.e.} a given density
$\rho=\frac{2}{d^2\sqrt{3}}$) is stable with respect to (harmonic)
vibrations if the frequencies of all normal modes are real. Using
Equation \ref{eq:disprel_hex}, and assuming $U'(d)<0$ (repulsive
potential), the stability condition becomes:
\begin{multline}\label{eq:hex_stabcond_long}
  \frac{U''(d)}{-U'(d)/d} \left( s_1^2+s_2^2+s_3^2+\eta s_0^2 \right)
  \geq \left( s_1^2+s_2^2+s_3^2-\eta s_0^2 \right),
\end{multline}
for all $\vec{k}$ in the Brillouin zone and $\eta=\pm 1$.  Noting that
$0\leq s_1^2+s_2^2+s_3^2-s_0^2 \leq s_1^2+s_2^2+s_3^2+s_0^2$, the
preceding condition can be rewritten as:
\begin{equation}\label{eq:func_anal}
  \frac{U''(d)}{-U'(d)/d}
  \geq
  \max_{\vec{k}\in\bz}\,
  \frac{s_1^2+s_2^2+s_3^2+s_0^2}{s_1^2+s_2^2+s_3^2-s_0^2}.
\end{equation}
An analysis of the function of $\vec{k}$ on the right--hand side of
Equation \ref{eq:func_anal} shows that the sought maximum is $3$, and
that it is achieved for all wavevectors lying along the $\vec{a}_1^*$,
$\vec{a}_2^*$, or $(\vec{a}_2^*-\vec{a}_1^*)$ axes of the Brillouin
zone (\textit{cf.} Figure \ref{fig:disprel_hex}).  The locus of the
maximum is thus compatible with the six--fold symmetry of the
reciprocal lattice.  The preceding inequality therefore reduces to:
\begin{equation}\label{eq:hex_stab_crit}
  \frac{U''(d)}{-U'(d)/d} \geq 3.
\end{equation}
For a given repulsive pair potential $U(x)$, Equation
\ref{eq:hex_stab_crit} determines the values of the density $\rho$ for
which the hexagonal Bravais lattice is stable with respect to harmonic
vibrations, in the nearest--neighbour approximation.
\section{Consequence for the two--dimensional crystal of composite
  bosons}
We now a consider a two--dimensional system of composite bosons
obtained in an ultracold mixture containing two different types of
Fermionic atoms. These composite bosons interact \textit{via} an
effective pair potential which is purely repulsive. An analytic
expression for this pair potential has been derived in the
Born--Oppenheimer approximation:
\begin{equation}\label{eq:U2d}
  U_\mathrm{2D}(R)=
  U_0\cdot 
  \left[
    \kappa_0 R \, \K_0(\kappa_0 R) \K_1(\kappa_0 R)
    -
    \K_0^2(\kappa_0 R)
  \right]
\end{equation}
where $\K_0$ and $\K_1$ are Bessel functions, $U_0$ is a constant, and
$\kappa_0^{-1}$ is the composite--boson molecular size.  This system
has been shown to exhibit a crystalline phase if the ratio of the two
different atomic masses is sufficiently large \cite{petrov:2007}.
The results presented in Sections \ref{sec:square} and \ref{sec:hex} 
provide a simple argument as to which
two--dimensional lattice, if any, the system crystallises into.

This ultracold system of composite bosons cannot be completely
described using classical mechanics. Indeed, the particles in the
system are not at rest, even at $T=0\,\mathrm{K}$: their positions
exhibit quantum zero--point fluctuations. However, if the system is in
a crystalline phase, this zero--point motion can be interpreted as a
vibration of the particles around the corresponding lattice
sites. Therefore, this (quantum) crystal can only be stable if the
corresponding crystal lattice is stable from a classical point of
view.

\begin{figure}
  \begin{center}
    \includegraphics[width=.6\linewidth]{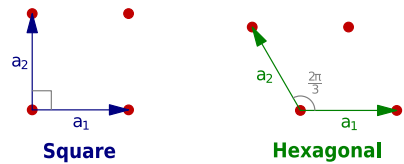}
    \caption{\label{fig:2d_lattices} Direct--lattice bases of the two
      2D Bravais lattices with high symmetry properties: the square
      lattice (four--fold symmetry) and the hexagonal lattice
      (six--fold symmetry).}
  \end{center}
\end{figure}

There are five types of two--dimensional Bravais lattices
\cite{kittel:1972}.  Among these, only two exhibit high symmetry
properties: the square lattice (four--fold symmetry) and the hexagonal
lattice (six--fold symmetry). The unit cells of both of these lattices
are represented in Figure \ref{fig:2d_lattices}.

All particles in the system are identical (they are all composite
bosons). It is therefore reasonable to assume that its crystal phase
will be highly symmetrical, and thus that the system crystallises in
either the square lattice or the hexagonal lattice. However, the
interaction between two composite bosons is characterised by a pair
potential which is repulsive for all relative distances. Therefore,
the results of Section \ref{sec:square_lattice_unstable} imply that
the square lattice is not a stable equilibrium position for this
system. Consequently, for values of the mass ratio and density leading
to a crystalline phase, the system will crystallise in a hexagonal
lattice. This prediction is confirmed by Quantum Monte Carlo
calculations \cite{petrov:2007}.

\begin{figure}
  \begin{center}
    \includegraphics[width=\linewidth]{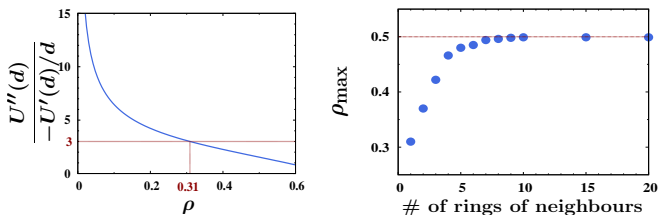}
    \caption{\label{fig:hex_stability} Stability of the hexagonal
      lattice in the case of the pair potential between composite
      bosons.  Left: plot of $\frac{U''(d)}{-U'(d)/d}$ as a function
      of the density $\rho=\frac{2}{d^2\sqrt{3}}$.  In the
      nearest--neighbour approximation, the hexagonal lattice is
      stable for $\frac{U''(d)}{-U'(d)/d}> 3$, \textit{i.e.} for
      $\rho<\rho^\mathrm{NN}_\mathrm{max}=0.31$.  Right: critical density
      $\rho_\mathrm{max}$ above which the hexagonal lattice is not
      stable, as a function of the number of rings of neighbours taken
      into account.  Starting from the eighth ring of neighbours the
      critical density saturates to $\rho_\mathrm{max}=0.499(1)$.  
      As in Figures \ref{fig:disprel_square} and \ref{fig:disprel_hex}, the
      unit of length is the composite--boson molecular size
      $\kappa_0^{-1}$.}
  \end{center}
\end{figure}

The range of densities for which the hexagonal lattice is stable is
determined, in the nearest--neighbour (NN) approximation, by the
criterion stated in Section \ref{sec:hex}. The relevant function
$\frac{U''(d)}{-U'(d)/d}$ is represented in Figure
\ref{fig:hex_stability} in the case of the pair potential \ref{eq:U2d}
(left--hand plot). The criterion for stability (Equation
\ref{eq:hex_stab_crit}) is satisfied for all densities lower than
$\rho_\mathrm{max}^\mathrm{NN}=0.31$.  For
$\rho>\rho_\mathrm{max}^\mathrm{NN}$, neither the square lattice nor
the hexagonal lattice are stable in the nearest--neighbour
approximation. Numerical calculations of $\rho_\mathrm{max}$ taking
into account farther rings of neighbours on a $100\times 100$
hexagonal lattice (Figure \ref{fig:hex_stability}, right--hand plot)
show that the corrections due to the next neighbours do not affect the
qualitative behaviour of the system: starting from the eighth ring of
neighbours, the critical density saturates to
$\rho_\mathrm{max}=0.499(1)$.  We therefore predict that, for densities
greater than $\rho_\mathrm{max}$, the system can exhibit no
crystalline phase: it is in a disordered phase regardless of the value
of the mass ratio.

Note that the numerical value of $\rho_\mathrm{max}$ that has just
been obtained must be considered with caution, since the expression of
the pair potential $U_{2D}$ (Equation \ref{eq:U2d}) that has been used
to derive it results from approximations that may not be strictly
valid in the present case. Nevertheless, it remains straightforward,
using our suggested procedure, to confirm the existence of a critical
density, and possibly refine its value given a more accurate pair
potential.
\section*{Discussion}
\paragraph*{Observability.}
The composite bosons are obtained in a trapped bipartite Fermi mixture
which has been cooled to degeneracy. The quasi--two--dimensional
regime can be reached by confining both types of atoms to the
antinodes of an optical lattice.  The crystalline or liquid phase of
the composite boson system may be characterised through
absorption--imaging techniques\cite{bourdel:2004,taglieber:2008}.
\paragraph*{Applicability.}
General theorems\cite{mermin:1968} have been stated, concerning a
specific --- albeit large --- class of pair potentials, which imply
that no crystalline order can be observed in infinite two-dimensional
systems. However, the composite-boson systems conceivable in
experiments are trapped, and hence finite-sized, systems, to which
these theorems do not apply, regardless of the specific shape of the
pair potential\cite{mermin:1968}. The experimental observation of a
two-dimensional crystalline phase of composite bosons will therefore
not contradict the theorems mentioned above.  Furthermore, hexagonal
lattices have already been observed in numerous other systems, such as
vortices in superconductors\cite{essmann:1967} and rotating
Bose-Einstein condensates\cite{aboshaeer:2002}, $\mathrm{C}_{60}$
molecules on a substrate\cite{yanagida:2000}, and colloidal
suspensions\cite{nakazawa:2006}.  In all four preceding cases, the
observed two-dimensional lattice is the hexagonal one, which
corresponds to our present prediction for the composite-boson system.

\section*{Conclusion}
The interactions of composite bosons in a two--dimensional ultracold
system are remarkable inasmuch as they are described by a
finite--ranged repulsive pair potential. In this context, we have
shown the square Bravais lattice to be unstable with respect to
harmonic vibrations, first through an analytic expression of its
dispersion relation derived using the nearest--neighbour
approximation, and then through numerical calculations taking into
account farther rings of neighbours.  Again using the
nearest--neighbour approximation, we have derived an analytic
expression of the dispersion relation for the hexagonal lattice. We
have stated a criterion determining the range of densities for which
this lattice is stable. In the particular case of the interaction
between composite bosons, this criterion yields a maximum density
above which no crystalline phase can be observed. Numerical
calculations have shown that taking into account farther rings of
neighbours does not qualitatively change the behaviour of the
system. We thus conclude that, for all values of the density and mass
ratio yielding a crystalline phase, the system of composite bosons
crystallises into the hexagonal lattice.

\begin{acknowledgments}
  The author wishes to thank Prof. G.V. Shlyapnikov and Dr. D.S. Petrov
  (LPTMS, Orsay) for initiating the present study, as well as for helpful
  suggestions. He also 
  acknowledges fruitful discussions with Y. Castin (LKB, ENS--Paris)
  and E. Lepage (DMA, ENS--Paris). LPTMS is research
  unit No. 8626 of CNRS and Universit\'e Paris--Sud.
\end{acknowledgments}

\end{document}